\documentclass[twocolumn,showpacs,preprintnumbers,amsmath,amssymb]{revtex4}

\usepackage{graphicx}
\usepackage{dcolumn}
\usepackage{bm}

\newcommand{\Sol}  {\textsc{sol}}

\newcommand{\Dms}  {\Delta m^2_\Sol}

\newcommand{\Dcq}  {\Delta\chi^2}

\def\roughly#1{\mathrel{\raise.3ex\hbox{$#1$\kern-.75em
\lower1ex\hbox{$\sim$}}}}
\def\lsim{\roughly<}

\begin{document}

\preprint{IFIC/03-49}

\title{Constraining the neutrino magnetic moment with 
  anti-neutrinos from the Sun}

\author{O. G. Miranda} 
\email{Omar.Miranda@fis.cinvestav.mx}
\affiliation{Departamento de F\'{\i}sica, Centro de Investigaci{\'o}n y de
  Estudios Avanzados del IPN\\ Apdo. Postal 14-740 07000 Mexico, DF, Mexico}


\author{T. I. Rashba}
\altaffiliation[On leave from ]{Institute of Terrestrial Magnetism,
    Ionosphere and Radio Wave Propagation of the Russian Academy of
    Sciences (IZMIRAN), 142190, Troitsk, Moscow, Russia}

\author{A. I. Rez}\email{rez@ific.uv.es}
\altaffiliation[On leave from ]{IZMIRAN}

\author{J. W. F. Valle}
\email{valle@ific.uv.es}
\homepage{http://www.ific.uv.es/~ahep/}
\affiliation{Instituto de F\'{\i}sica Corpuscular -- C.S.I.C.,
  Universitat de Val{\`e}ncia \\
  Edificio Institutos, Apt.\ 22085, E--46071 Val{\`e}ncia, Spain}

\date{\today}

\begin{abstract}
  We discuss the impact of different solar neutrino data on the
  spin-flavor-precession (SFP) mechanism of neutrino conversion. We
  find that, although detailed solar rates and spectra allow the SFP
  solution as a sub-leading effect, the recent KamLAND constraint on
  the solar antineutrino flux places stronger constraints to this
  mechanism. Moreover, we show that for the case of random magnetic
  fields inside the Sun, one obtains a more stringent constraint on the
  neutrino magnetic moment down to the level of $\mu_\nu \lsim {\rm
    few} \times 10^{-12}\mu_B$, similar to bounds obtained from
  star cooling.
\end{abstract}

\pacs{26.65.+t Solar neutrinos 96.60.Jw Solar interior 13.15.+g
  Neutrino interactions
14.60.Pq  Neutrino mass and mixing }
\maketitle


The latest KamLAND result~\cite{:2003gg} greatly improves the expected
sensitivity limit on a possible antineutrino component in the solar
flux from 0.1 \%~\cite{Kamland:proposal} of the solar boron $ \nu_e $
flux to $ 2.8\times 10^{-2}$ \% at the 90 \% C.L., about 30 times
better than the recent Super-K limit~\cite{Gando:2002ub}.

The presence of anti-neutrinos in the solar flux may signal the
existence of SFP conversions induced by non-vanishing neutrino
transition magnetic moments~\cite{Schechter:1981hw,Akhmedov:uk} or,
alternatively, neutrino decays in models with spontaneous violation of
lepton number~\cite{Schechter:1981cv}.  Here we focus on the more
likely case of anti-neutrinos produced by SFP conversions.  Several
solar magnetic field models are possible, characterized by different
assumptions pertaining to their magnitude, location and typical
scales~\cite{Kutvitskii,others,Friedland}.

\begin{figure*}
\includegraphics[width=0.75\textwidth,height=3cm]{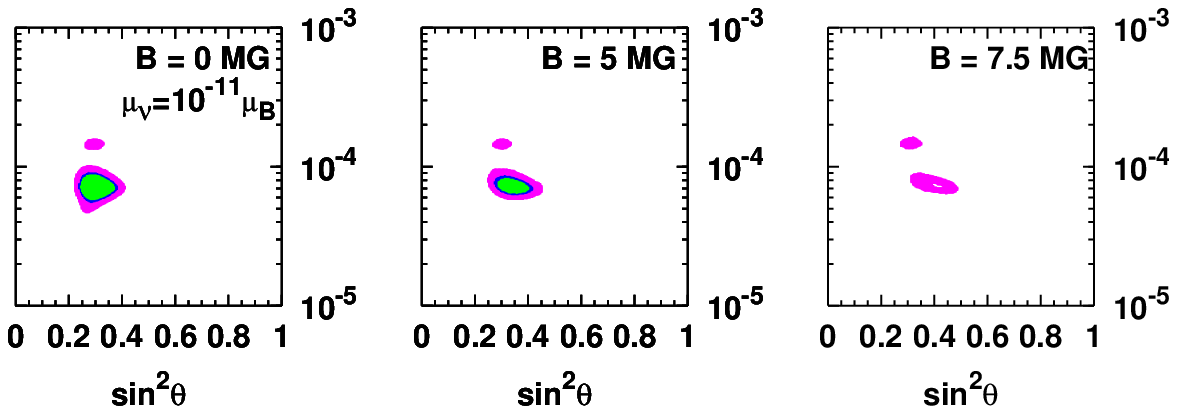}
\includegraphics[width=0.7\textwidth,height=3cm]{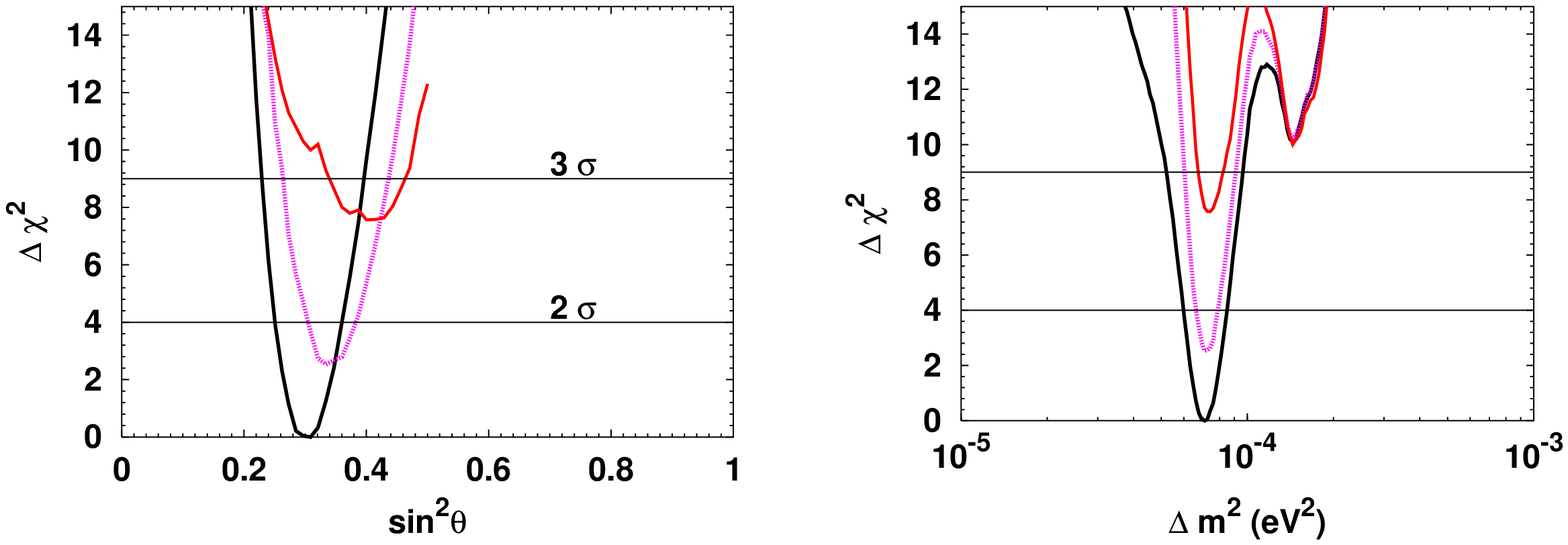}
\caption{\label{fig:panels} Solar neutrino oscillation parameters
  for zero (upper left panel) and non-zero (upper right panels)
  self-consistent CZ magnetic field used in Ref.~\cite{Kutvitskii}. Lower
  panels give $\Dcq$ profiles with respect to $\Dms$ and $\sin^2
  \theta_\Sol$.}
\end{figure*}

Previous to the latest KamLAND limit on the solar anti-neutrino flux,
the first KamLAND evidence of reactor anti-neutrino
disappearance~\cite{Eguchi:2002dm} had already excluded SFP scenarios
as solutions to the solar neutrino problem~\cite{Barranco:2002te}.
However, even with the recent SNO salt results~\cite{Ahmed:2003kj}
which confirm the simplest three-neutrino oscillation
picture~\cite{Maltoni:2003da}, a neutrino magnetic moment could still
play a role as a sub-leading effect. In order to illustrate this, we
have performed a $\chi^2$ analysis taking into account all the solar
experimental data plus the KamLAND disappearance data for the same
self-consistent profile for the magnetic field in the convective zone
used in previous analyses \cite{Miranda:2000bi}.  The results are
shown in fig.~\ref{fig:panels}, were we have taken different values
for the product $\mu_\nu B$ (neutrino magnetic moment in units of
$10^{-11}\mu_B$ and maximum magnetic field value in convective zone in
units of MG) and plotted the allowed regions at 90~\%, 95~\% and 99~\%
C. L.

One can see how the allowed LMA-MSW region of oscillation parameters
(upper left panel) is modified in the presence of SFP conversions
(upper right panels) in Fig.~\ref{fig:panels}.  The lower panels in
this figure give the $\Dcq$ profiles with respect to $\Dms$ and
$\sin^2 \theta_\Sol$, from where one can determine the corresponding
allowed ranges. Given the current laboratory best limit on the
neutrino magnetic moment $\mu_{\bar \nu_e} < 1.0\times10^{-10}\mu_B
$~\cite{Daraktchieva:2003dr,Grimus:2002vb}, and the fact that magnetic
field amplitude inside the convective zone should not exceed
0.1-0.3~MG~\cite{Moreno} one can see that, although limited, there
is room left for sub-dominant SFP conversions.

We now turn to the recent KamLAND result on the $\bar{\nu_e}$~'s from
the Sun. The Collaboration has reported that the antineutrino flux
$\Phi_{\bar{\nu_e}}$ is less than $3.7\times
10^{2}$~cm~$^{-2}$~s~$^{-1}$ at 90 \% C. L. which corresponds to a
0.028 \% of the solar $^8$~B $\nu_e$ flux (in the energy window
between 8.3~MeV and 14.8~MeV). It is this last result that will
substantially limit the possibility of subleading SFP component in the
neutrino conversion mechanism, establishing the robustness of the
oscillation hypothesis. 

Within our generalized picture (LMA-MSW+SFP), after the MSW
$\nu_e\to\nu_\mu$ conversion takes place in the central region of the
Sun, $\bar{\nu_e}$'s are produced due to the magnetic moment
conversion $\nu_\mu\to\bar{\nu_e}$.  To analyse quantitatively the
restrictions imposed by this result we consider three models for the
solar magnetic field:
\begin{enumerate}
\item regular magnetic fields, both in the convective
  (CZ)~\cite{Kutvitskii} and radiative (RZ)~\cite{Friedland} zones of
  the Sun
\item convective zone random magnetic fields~\cite{Bykov:1998gv}
\end{enumerate}

For definiteness we consider the simplest aproximate two-neutrino
picture, which is justified in view of the stringent limits that
follow mainly from reactor neutrino experiments~\cite{Maltoni:2003da}.
From the resulting $4\times 4$ form of the neutrino evolution
equation~\cite{Schechter:1981hw,Akhmedov:uk}, we compute the expected
$\bar{\nu_e}$ yield for each magnetic field model. For the case of CZ
magnetic fields, this $4\times 4$ form of the neutrino evolution
further decouples into LMA-MSW conversions deep in the Sun followed by
(approximate) vacuum SFP conversions~\cite{Schechter:1981hw} inside
the CZ. In contrast, for the RZ case there is no decoupling of the
neutrino evolution, due to the large strength of the magnetic
field~\footnote{Although possible, this model is not favored on
  physical grounds.}.

\begin{figure*}
\includegraphics[width=0.7\textwidth,height=4cm]{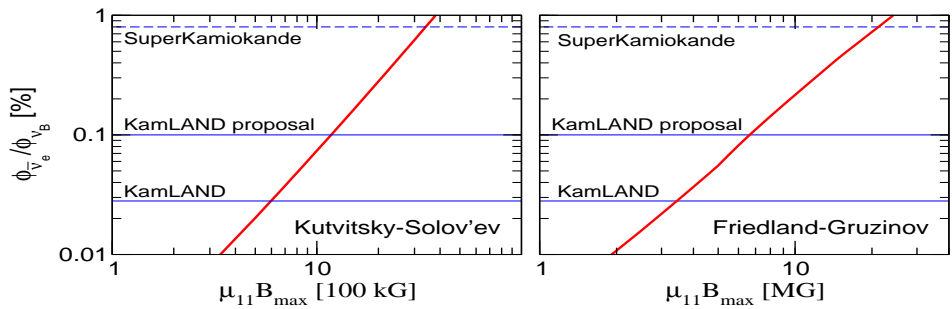}
\caption{\label{fig:reg} Bounds on $\mu_{\nu} B$ for regular magnetic
  field models, Kutvitsky-Solov'ev (left) and Friedland Gruzinov (right). The
  horizontal lines indicate the bounds on solar anti-neutrinos from
  Super-K and KamLAND.}
\end{figure*}

In order to obtain a conservative bound on $\mu_\nu B$ we have
calculated the minimum $\bar{\nu_e}$ yield as the oscillation
parameters vary within the acceptable range (at the 90~\% C.L.) in the
absence of magnetic field (pure LMA-MSW case).  Such minimum predicted
anti-neutrino fluxes for the case of regular field profiles are shown
in the curves depicted in Fig.~\ref{fig:reg}, while the recent KamLAND
limit is indicated by the lower horizontal lines. For comparison, we
present the previous Super-K bound, indicated by the upper horizontal
lines.  Note that in both cases of regular field profile we can only
constrain the product $\mu_{\nu} B$, as opposed to the intrinsic
neutrino magnetic moment~\footnote{For constraints on the intrinsic
  neutrino magnetic moment see Ref.~\cite{Grimus:2002vb}. }.
~\cite{Grimus:2002vb}.

A more interesting picture for SFP scenario is obtained if we consider
the case of turbulent random magnetic fields inside the Sun. As we
will see, the reason for this is twofold: (i) we will be able to fix,
to some extent, the dependence on the magnetic field, in contrast to
previous analysis of the random magnetic field presented in
Ref.~\cite{Bykov:1998gv,Torrente-Lujan:1998sy}, where an extra parameter
$L_0$ appeared, characterizing the scale of the random magnetic field
cells, and (ii) this model leads to more stringent limits on the
neutrino magnetic moment by itself.

In the following we give a brief discussion of our new approach to the
problem (a more detailed analysis will be published
elsewhere~\cite{Inprep}).  The mean magnetic field value over the
solar disc is of the order of 1 G and magnetic field strength in the
solar spots reaches 1 kG. It is commonly accepted that fields measured
at the solar surface are substantially weaker than those near the
bottom of the convective zone (CZ) where they are supposed to be
generated by a dynamo mechanism~\cite{Yoshimura,Sokoloff}. In dynamo
theory, the mean magnetic field is accompanied by a small-scale random
magnetic field, which is not directly traced by sunspots or any other
tracers of solar activity.  By small scales we denote a typical scale
of about the solar granule size ($\sim$ 1000 km)~\cite{Stix} where the
rms magnetic field amplitudes (generated either as a by-product of a
large-scale dynamo mechanism or directly by a small-scale dynamo) in
the range of 50-100 kG are reasonable.

For LMA oscillations the MSW conversion occurs well below the CZ
resulting in a coherent mixture of two neutrino flavours at the bottom
of the convective zone.  When neutrinos cross the randomly fluctuating
magnetic field of the CZ it is expected that $\nu_{eL}$ will convert
to ${\bar\nu_{\mu R}}$ and $\nu_{\mu L}$ will convert to $\bar \nu_{e
  R}$ as a result of the CZ magnetic field. For the case of random CZ
magnetic fields these populations will tend to equilibrate.  However,
given the laboratory upper limits on $\mu_{\nu}$ and the finite depth
of the CZ, this equilibration is not achieved. The relaxation can be
viewed as a small-amplitude random walk of the neutrino polarization
vector leading to a small appearance of electron and muon
antineutrinos at the solar surface ({\em cf.}  \cite{Loeb:1989dr} for
the vacuum case, and also discussion in~\cite{Bykov:1998gv}). In what
follows we describe the main features of this calculation and give the
main result.

Because of the randomness of the underlying magnetic fields, the spin
flavour evolution looses coherence, {\em i.e.}, instead of evolving
neutrino wave functions over the whole CZ and then obtaining the final
probabilities, one has to compute the probabilities in each
correlation cell and afterwards add them.  As a result, independently
of the random magnetic field model, the appearance of antineutrinos is
proportional to the relevant Fourier harmonic of the transverse
two-point magnetic correlation function, with space period equal to
the LMA-MSW neutrino oscillation length ({\em cf.}~\cite{Loeb:1989dr}
and discussion in~\cite{Bamert:1997jj} for the MSW matter noise).

We will assume for definiteness that random magnetic field evolution
is due to the highly developed steady-state MHD turbulence treated
within the Kolmogorov scaling theory~\cite{Kolmogorov,Landau}. The
magnetic Reynolds number within the CZ is very large, $R_m \sim 10^8$
\cite{Sokoloff}, and the corresponding inertial range is very
wide,~$l_{\rm diss}< l<L_{\rm max}$ where the outer scale is supposed
to be $L_{\rm max} \simeq 1000$~km, about the solar granule size, and
~ $l_{\rm diss} = L_{\rm max} R_m^{-3/4} \sim 1~$~m. The rms magnetic
field $b_l$ is assumed to scale as $b_l \sim l^{1/3}$.  This implies
that, after fixing the maximum field amplitude at the outer scale
$L_{\rm max}$ it is straightforward to obtain the rms field at the
neutrino oscillation scale, which is about several hundreds kilometers
for LMA-MSW case, well within the inertial range where scaling
arguments are valid~\cite{Inprep}. Here we show only the final result
for neutrino mass eigenstate probabilities at the surface of the Sun:
\begin{equation}
|\nu_{1L}|^2_{R_\odot} = P_1(1-\eta), \qquad  
|\nu_{2R}|^2_{R_\odot} = P_1 \eta~,
\end{equation}
\begin{equation}
|\nu_{2L}|^2_{R_\odot} = P_2(1-\eta), \qquad  
|\nu_{1R}|^2_{R_\odot} = P_2 \eta~,
\label{EqnAtSunSurface}
\end{equation}
(note that these obey unitarity) where the common factor $\eta$ is
given by
\begin{equation}
\label{Feta}
\eta \simeq 0.3~ 
\frac{{\mu_{\nu}}^2 \bar{b}_{\rm max}^2}{{\delta^2 }} 
\frac{\delta \cdot L }
{(\delta \cdot L_{\rm{max}})^{2/3}}S^2~.
\end{equation}

Here $P_i = P_{iL}=|\nu_{iL}(r=0.7R_\odot)|^2$ are the probabilities
that solar neutrinos reach the bottom of the CZ in a given mass state,
$L$ is the CZ width, $\delta = \Delta m^2/4E$, and $S^2$ is a rms
magnetic field profile shape factor
\begin{equation}
S^2=\frac{1}{L}\int_0^L dz \frac{{\bar b^2}(z)}{{\bar b_{max}}^2}.
\label{S}
\end{equation}
Note that $S=1$ for constant shape of rms field and of the order of
unity for other profiles, e.g.  $S=0.579$ for the ``smooth'' profile
of~\cite{Bykov:1998gv}, and $S=0.577$ for a triangle profile.

The different factors in Eq. \ref{Feta} can be explained if we keep in
mind that $\delta$ is inversely proportional to the neutrino
oscillation length, $\delta= \pi / \lambda_{osc}$~. The ratio
$\bar{b}_{\rm max}^2/(\delta \cdot
L_{\rm{max}})^{2/3}=\bar{b}_{\lambda_{\rm osc}}^2$ is the squared rms
field at the scale $l=1/\delta\sim\lambda_{\rm osc}$, and the ratio
$\mu^2_\nu \bar{b}_{\lambda_{\rm osc}}^2 / \delta^2$ determines the
fraction of neutrinos which experience (on average) spin-flavour
conversion to the antineutrino states in a correlation cell of length
$l\sim \lambda_{\rm osc}$. Finally $\delta\cdot L\sim L/\lambda_{\rm
  osc}$ is the number of correlation cells along the neutrino
trajectory and the factor $S^2$, as already mentioned, accounts for
the specific shape of the rms field profile. (The numerical factor 0.3
comes from the integration along the trajectory).

We can rewrite Eq. \ref{Feta} in normalized units as
\begin{equation}
\label{FetaRenormalized}
\eta \sim 3\times 10^{-3} 
\mu_{11}^2 \varepsilon^2 S^2 
\left(\frac{7\times 10^{-5} eV^2}{\Delta m^2} \right)^{5/3}
\left(\frac{E}{10 {\rm MeV}}\right)^{5/3}
\end{equation}
where $\mu_{11}$ is the magnetic moment in units of $10^{-11}\mu_B$,
and the ratio $\varepsilon= (b/100~{\rm kG})(L_{{\rm max}}/1000~{\rm
  km})^{1/3}$.

Although the ratio $\varepsilon$ is not known precisely, we can obtain
a good estimate assuming the equipartition between kinetic energy of
hydrodynamic pulsations and the rms magnetic energy at the upper (most
energetic) scale $L_{max}$~\cite{Sokoloff}
\begin{equation}
\frac{\rho \bar{v^2}}{2}\sim \frac{\bar{b}_{\rm max}^2}{8\pi}
\end{equation}
Taking $v\sim 3\times 10^4 {\rm cm}/{\rm s}$, $\rho \sim 1g/ {\rm
  cm}^3$ we obtain $b_{\rm max}\sim100$~kG. In what follows we will
consider the range $b_{\rm max}\sim 50-100$~kG.

\begin{figure}
\includegraphics[height=4cm,width=0.35\textwidth]{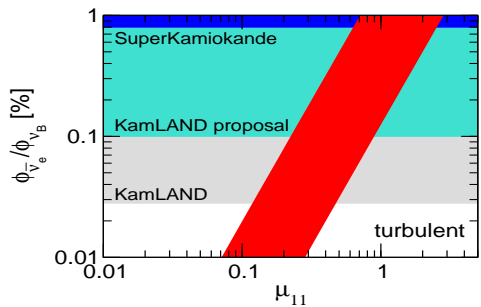}
\caption{\label{fig:turbulent} Bounds on $\mu_{\nu} $ for the turbulent
  magnetic field model described in the text. The horizontal lines
  indicate the bounds on solar anti-neutrinos from Super-K and
  KamLAND.}
\end{figure}

This specific behaviour implies, on physical grounds, that
$\varepsilon$ is not allowed to vary substantially, $0.5 <\varepsilon
< 1$, with our assumptions. As mentioned, for different profiles we
found that the parameter $S^2$ lies in the range between 0.5 and 1.
With this in mind, we can see that the product $k=\varepsilon S$ lies
in the interval $0.25<k<1$.  Since the overall $\nu_e \to \bar{\nu_e}$
conversion probability depends linearly on $\eta$, it follows that, by
comparing the resulting $\bar{\nu_e}$ yield to the recent KamLAND
bound one can constrain the value of $\mu_\nu$ to within a factor 4.
More precisely, we have computed the solar anti-neutrino yield for
this random magnetic field model for all allowed values of neutrino
oscillation parameters in the LMA-MSW region. Our results are shown in
fig.~\ref{fig:turbulent} as a function of $\mu_\nu$, where the width
of the band corresponds to the freedom in choosing the parameter $k$.
It is easy to see that, indeed, the constraint we obtain is better
than those that hold for the case of regular magnetic fields.
Moreover, from our estimate for the minimum value of $k$, we can
obtain an upper bound for $\mu_{\nu}$ which lies in the range $\mu_\nu
\leq 1.5 \times 10^{-12}\mu_B$ to $\mu_\nu\leq 5 \times
10^{-12}\mu_B$.
We have obtained this result for the Kolmogorov theory. However, it
can be easily generalized for arbitrary power-law exponent $p$ (for
Kolmogorov theory $p = 5/3$). We have checked that our bound on
$\mu_\nu$ is rather robust with respect to possible changes of the
power-law exponent $p$ within the interval from 1 to 2~\cite{Inprep}.

In short, we have discussed the impact of recent solar neutrino data
on the spin-flavor-precession (SFP) mechanism of neutrino conversion.
The recent KamLAND bound on the solar anti-neutrino yield leads to a
constraint on the product of Majorana neutrino transition magnetic
moment and some magnetic field strength in different models for the
solar magnetic field.  For the case of random magnetic fields inside
the Sun, one can obtain a direct constraint on the intrinsic neutrino
magnetic moment of $\mu_\nu \lsim {\rm few} \times 10^{-12}\mu_B$,
similar to bounds obtained from star cooling.  Comparing these
constraints with Fig.~\ref{fig:panels} one sees that the robustness of
the oscillation interpretation of current solar neutrino data against
possible magnetic-field-induced transitions is firmly established.

We thank V. B. Semikoz and D. D. Sokoloff for useful discussions.
This work was supported by Spanish grant BFM2002-00345, by European
RTN network HPRN-CT-2000-00148, by European Science Foundation network
grant N.~86, by INTAS YSF grant 2001/2-148 and MECD grant SB2000-0464
(TIR). TIR and AIR were partially supported by the Presidium RAS and
CSIC-RAS grants.  OGM was supported by CONACyT-Mexico and SNI.

\end{document}